\documentclass[aps,prl,reprint]{revtex4-2}
\usepackage[utf8]{inputenc}
\usepackage{graphicx}
\usepackage{amsfonts,amsthm}
\usepackage{amsmath,amssymb}
\usepackage{bm}
\usepackage[usenames,dvipsnames]{xcolor}

\newcommand{\lr}[1]{ \langle #1 \rangle}

\newcommand{\doublet}[2]{ \left( \begin{array}{c}#1 \\ #2 \end{array}\right) }

\newcommand{\dk}{ \widetilde{dk} }

\newcommand{\bb}{{\bf b}}

\newcommand{\bk}{{\bf k}}
\newcommand{\bK}{{\bf K}}

\newcommand{\bp}{{\bf p}}

\newcommand{\br}{{\bf r}}

\def\lsim{\mathrel{\rlap{\lower4pt\hbox{$\sim$}}
	\raise1pt\hbox{$<$}}}         
\def\gsim{\mathrel{\rlap{\lower4pt\hbox{$\sim$}}
	\raise1pt\hbox{$>$}}}         

\graphicspath{{./figures/}}

\begin{document}
\title{Unambiguous detection of high energy vortex states via the superkick effect}

\author{Zhengjiang Li$^{1}$} 
\email[These authors contributed equally to this work.]{}
\author{Shiyu Liu$^{2}$} 
\email[These authors contributed equally to this work.]{}
\author{Bei Liu$^{1}$}
\email[These authors contributed equally to this work.]{}
\author{Liangliang Ji$^{2}$}
\email{jill@siom.ac.cn}
\author{Igor P. Ivanov$^{1}$}
\email{ivanov@mail.sysu.edu.cn}

\affiliation{$^{1}$School of Physics and Astronomy, Sun Yat-sen University, Zhuhai 519082, China\\
$^{2}$State Key Laboratory of High Field Laser Physics, Shanghai Institute of Optics and Fine Mechanics (SIOM), Chinese Academy of Sciences, Shanghai 201800, China}

\date{\today}

\begin{abstract}
Vortex states of photons, electrons, and other particles 
are freely propagating wave packets
with helicoidal wave fronts winding around the axis of a phase vortex. 
A particle prepared in a vortex state carries
a non-zero orbital angular momentum projection on the propagation direction,
a quantum number that has never been exploited in experimental particle and nuclear physics. 
Low-energy vortex photons, electrons, neutrons, and helium atoms 
have been demonstrated in experiment and found numerous applications,
and there exist proposals of boosting them to higher energies.
However, verification that a high energy particle is indeed in a vortex state
will be a major challenge, since the low energy techniques become impractical at higher energies.
Here, we propose a new diagnostic method based of the so-called superkick effect,
which can unambiguously detect the presence of a phase vortex. 
A proof-of-principle experiment with vortex electrons can be done with existing technology,
and its realization will also constitute the first observation of the superkick effect.
\end{abstract}

\maketitle


The structure and dynamics of composite systems, from atoms to hadrons, 
is an immensely rich research field. 
They are usually probed in collision experiments, such as atomic spectroscopy \cite{Kakkar:2015}
or deep inelastic scattering of energetic electrons from protons \cite{Roberts:1990ww},
and polarization often gives access to additional insights
on the composite system structure \cite{Aidala:2012mv}.

A decade ago it was realized that yet another, previously unexplored degree of freedom 
can further expand experimental capabilities of particle collisions
in atomic, nuclear, and particle physics \cite{Bliokh:2007ec,Jentschura:2010ap,Ivanov:2011kk,Karlovets:2012eu}. 
This new degree of freedom is the non-zero orbital angular momentum (OAM) projection 
on the propagation direction. 
Any particle, be it a photon, an electron, or a composite object such as a hadron, 
can be endowed with a well-controlled OAM by preparing it as a wave packet with helicoidal wave fronts,
known as a vortex, or twisted, state. This wave packet possesses zero intensity 
on the phase singularity axis and the phase factor $\psi(\br) \propto e^{i \ell \varphi_r}$
in its vicinity, where $\varphi_r$ is the azimuthal angle in the transverse plane. 
For a scalar particle, 
it leads to a well-defined OAM $L_z = \hbar \ell$; 
for particles with spin, solutions can be constructed as eigenstates of the helicity 
and the total angular momentum projection operators, \cite{Bliokh:2017uvr,Lloyd:2017,Knyazev-Serbo:2018,Ivanov:2022jzh}.

Vortex photons are known since decades \cite{Allen:1992zz} and have found numerous applications \cite{Torres-applications}.
Vortex X rays have also been produced \cite{Nature-Phot-2019},
and ideas on generating hard gamma photons in the MeV and GeV ranges have been outlined \cite{Jentschura:2010ap,Taira:2017,Chen:2019}.
Following the proposal of \cite{Bliokh:2007ec}, three groups produced vortex electrons
with the kinetic energy up to 300~keV \cite{Uchida:2010,Verbeeck:2010,McMorran:2011},
with immediate applications in nanoresearch \cite{Verbeeck:2011-atomic}.
Recently, cold neutrons \cite{Sarenac:2022} and slow helium atoms \cite{luski2021vortex} were 
also put in vortex states.
Various schemes of producing high energy vortex states of electrons and ions
or of accelerating lower energy vortex particles to higher energies are being discussed \cite{workshop}. 

Once a vortex state is produced, one must verify that it indeed carries a non-zero OAM.
A hallmark feature of a vortex state is its ring-shaped intensity profile in the transverse plane,
which, for high-energy vortex electrons, can be revealed via inverse Compton scattering \cite{Seipt:2014bxa,Bu:2023fdj},
electron scattering on target atoms \cite{Serbo:2015kia,Karlovets:2016uhb},
or Vavilov-Cherenkov radiation \cite{Kaminer:2014iia,Ivanov:2016xab,Chaikovskaia:2023xtr}.
Although indicative of a vortex state, the ring-shaped intensity distribution alone
cannot distinguish a phase vortex from a non-vortex ring-shaped wave function. 
Alternatively, transition radiation by the OAM-induced magnetic moment of the vortex electron 
was proposed in \cite{Ivanov:2013eqa} as another method to verify the vortex state. 
However, detecting this effect seems feasible only for very large OAM values.

Vortex state can be unambiguously identified only if one directly probes 
the phase vortex and detects its winding number $\ell$.
At lower energies, this is routinely done with fork diffraction gratings \cite{Bliokh:2017uvr,Lloyd:2017,Knyazev-Serbo:2018}.
At higher energies, this method becomes impractical due to the very short
de Broglie wavelength and high penetrating power of energetic particles.
Interference of the vortex state with a reference plane wave, another diagnostic tool used at lower energies \cite{Uchida:2010}, 
requires transverse coherence lengths of the order of a micron, 
which has not yet been demonstrated for ultrarelativistic electrons in accelerators and storage rings \cite{Karlovets:2020odl}.
Finally, a generalized measurement procedure was suggested in \cite{Karlovets:2022evc,Karlovets:2022mhb}
to reveal the phase structure of the final wave function.
However, it relies on a new class of wavefront-sensitive detectors 
which are yet to be demonstrated at high energies.

In this Letter, we propose a method which can unmistakably reveal the presence of a phase vortex
in a high-energy wave packet, which is free of the drawbacks of the previous suggestions.
The method relies on elastic scattering of a vortex state with a compact non-vortex probe wave packet
and reveals the existence of the phase vortex through the peculiar feature in the total final momentum 
angular distribution known as the superkick effect.
The vortex beam is not blocked by any target, 
and the final state particles can be detected with traditional detectors.
The method is applicable at low and high energies and is operational 
even if the detector cannot resolve the vortex ring structure through the direct beam imaging.

We will illustrate the idea with M\o{}ller (elastic $e^-e^-$) scattering, 
but the effect is purely kinematical and exists for other particles. 
A proof-of-principle experiment is feasible with existing technology and will, 
simultaneously, demonstrate the superkick effect for the first time.


{\em M\o{}ller scattering of wave packets.}---We begin with the plane-wave (PW) M\o{}ller scattering,
in which two initial electrons with the four-momenta $k_1^\mu = (E_1,\bk_1)$, $k_2^\mu = (E_2,\bk_2)$
scatter elastically into the two final electrons with the four-momenta $k_3^\mu = (E_3,\bk_3)$, 
$k_4^\mu = (E_4,\bk_4)$. Throughout the paper, we use the relativistic Lorentz-Heaviside units 
$\hbar = c = 1$, $e^2 = 4\pi\alpha_{em}$, 
and denote three-vectors with bold symbols, adding the subscript $\perp$ for transverse vectors. 
We will also use the shorthand notation for the Lorentz-invariant momentum space measure
$\dk \equiv d^3k/[2E(k) (2\pi)^3]$.

The PW $S$-matrix element is
\begin{equation}
	S_{PW} = i(2\pi)^4\delta^{(4)}(k_1+k_2 - k_3-k_4) \frac{{\cal M} \cdot N_{PW}^4}{\sqrt{16 E_1 E_2 E_3 E_4}}\,,
	\label{SPW}
\end{equation}
where $E(k) = \sqrt{k^2 + m_e^2}$, $m_e$ being the electron mass.
The invariant amplitude ${\cal M}$ written in general kinematics derived in \cite{Ivanov:2016oue}. 
The normalization factor $N_{PW} = 1/\sqrt{V}$ corresponds to one particle per large volume $V$.
Since the initial $\bk_1$ and $\bk_2$ are fixed, the total final state energy $E_3 + E_4 = E_f = E_1 + E_2$
and momentum $\bk_3 + \bk_4 \equiv \bK = \bk_1 + \bk_2$ 
is also fixed. 

Beyond PW collisions, one can represent the initial particles 
as momentum space wave packets $\phi_1(\bk_1)$ and $\phi_2(\bk_2)$ normalized as
$\int \dk |\phi_i(\bk)|^2 = 1$
and apply the scattering theory of arbitrarily shaped beams 
\cite{Kotkin:1992bj,Karlovets:2018iww,Karlovets:2020odl}.
We focus on the head-on collision of wave packets which are centered around their respective 
average momenta $\bp_1 = (0,0,p_2)$ and $\bp_2 = (0,0,p_2)$, $p_2 < 0$.
Assuming that the momentum spread inside each wave packet is much smaller than $p_i$,
which allows us to treat the process in the paraxial approximation.
The differential probability of wave packet scattering into a PW final state is
\begin{equation}
	dW = (2\pi)^8 |{\cal I}|^2 \dk_3 \dk_4\,.
	\label{dsigma-WP}
\end{equation}
where
\begin{equation}
	{\cal I} = \int\dk_1 \dk_2 \,\phi_1(\bk_1)\,\phi_2(\bk_2)
	\delta^{(4)}(k_1+k_2 - K)\,\cdot{\cal M}\,,\label{cal-I}
\end{equation}
where $K^\mu = (E_f, \bK)$.

As the initial wave packets are no longer momentum eigenstates, 
the total final momentum $\bK$ is not fixed, and one can explore the differential cross section
as a function not only of $\bk_3$ but also of $\bK$.

\begin{figure}[!h]
	\centering
	\includegraphics[width=0.4\textwidth]{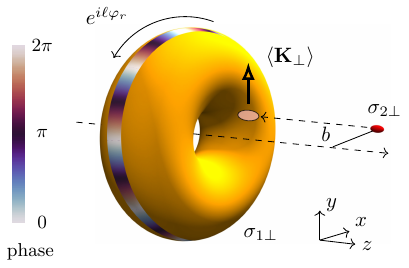}
	\caption{A broad LG and a compact Gaussian wave packets of the transverse sizes 
		of $\sigma_{1\perp}$ and $\sigma_{2\perp}$ collide at a non-zero impact parameter $b$ along axis $x$.
		The donut shape with a colored stripe shows the region of high intensity and the phase of the LG wave packet. 
		In the overlap region, shown by the pink ellipse, the azimuthally varying phase induces a total transverse momentum 
		$\lr{\bK_\perp}$ orthogonal to $\bb_\perp$.
	}
	\label{fig-collision}
\end{figure}

{\em The superkick effect.}---Let us apply this scheme to a head-on collision
of a vortex Laguerre-Gaussian (LG) wave packet 
with a compact Gaussian wave packet, Fig.~\ref{fig-collision}. 
In coordinate space, the two wave packet centroids follow parallel lines separated 
by a transverse impact parameter $\bb_\perp$. 
We choose the LG wave packet in the principal radial mode described by 
\begin{eqnarray}
	\phi_1(\bk_1) &\propto & \frac{(\sigma_{1\perp}k_{1\perp})^{\ell}}{\sqrt{\ell!}}\, e^{i\ell\varphi_k}\,
	\, e^{ -k_{1\perp}^2 \sigma_{1\perp}^2/2}\, e^{-(k_{1z}-p_{1})^2 \sigma_{1z}^2/2}\nonumber
\end{eqnarray}
and the Gaussian wave packet with
\begin{eqnarray}
	\phi_2(\bk_2) \propto e^{-i \bb_\perp \bk_{2\perp}}
	\,e^{ -k_{2\perp}^2 \sigma_{2\perp}^2/2}\, e^{-(k_{2z}-p_{2})^2 \sigma_{2z}^2/2}\nonumber\,.
\end{eqnarray}
The transverse $\sigma_{1\perp}$, $\sigma_{2\perp}$ and longitudinal $\sigma_{1z}$, $\sigma_{2z}$
extents can be chosen independently \footnote{See Supplemental Material, which includes Refs.~\cite{Sheremet:2024jky,Landau4}, 
	for additional details on wave packet definitions, helicity amplitude calculations,
	and an order of magnitude estimate of the scattering probability.}.

The problem possesses three transverse spatial scales:
the widths of the two wave packets at maximal focusing $\sigma_{1\perp}$
and $\sigma_{2\perp}$ and the impact parameter $b \equiv |\bb_\perp|$.
We focus on the regime $b\sim \sigma_{1\perp} \gg \sigma_{2\perp}$,
when a compact Gaussian wave packet probes the wider LG ring.
By detecting the scattered electrons and measuring $\bk_3$ and $\bk_4$,
one can determine the total final transverse momentum $\bK_\perp$ on the event-by-event basis.
Its average value $\lr{\bK_\perp}$ is non-zero and, remarkably, has a component orthogonal to $\bb_\perp$.
What is surprising is that $\lr{\bK_\perp}$ can be much larger than $\sqrt{\ell}/\sigma_{1\perp}$,
the typical transverse momentum of the LG state.
This unexpectedly large total transverse momentum of the scattered final state 
was dubbed in \cite{barnett2013superkick} as the superkick.
Semiclassically, it originates from the rapidly varying LG phase factor $e^{i\ell \varphi_r}$
near the phase singularity axis. As we go around the axis at a distance $b$,
this phase factor accumulates the phase change $2\pi \ell$ along the circle 
of circumference $2\pi b$. This rapid phase variation
induces the azimuthal component of the local momentum distribution,
$p_\varphi = \ell/b$, which can become arbitrarily large as $b \to 0$.
The finite $\sigma_{2\perp}$ tames the $1/b$ dependence at about $b \sim \sigma_{2\perp}$ 
but still allows for the peak value of $\lr{\bK_\perp}$ much larger than 
the transverse momentum of the LG state
\cite{Ivanov:2022sco,Liu:2022nfq}.

The superkick effect, initially proposed for twisted light absorbed by a trapped atom \cite{barnett2013superkick}, 
was recently predicted for hadronic photoproduction in \cite{Afanasev:2020nur,Afanasev:2021fda}
and investigated in detail for a generic two particle scattering \cite{Ivanov:2022sco,Liu:2022nfq}.
No experimental observation has been reported yet, 
the main challenge being the need to detect the small momentum transfer to a trapped atom
upon absorption of a vortex photon. The scheme we propose readily overcomes this challenge
and, as a by-product, can lead to the first experimental demonstration of the superkick effect.


{\em The proposal.}---The superkick effect is an unambiguous signature of a phase vortex.
To prove that a high energy electron is in a vortex state,
we propose to run an elastic $ee$ scattering experiment at a fixed $b$. 
When electrons scatter with a sufficiently large momentum transfer, 
the detectors installed close to the beam axis capture them in coincidence and
measure their deflection angles, from which their transverse momenta $\bk_{3\perp}$ and $\bk_{4\perp}$
and the total final transverse momentum $\bK_\perp = \bk_{3\perp} + \bk_{4\perp}$ can be computed. 
This measurement can be done on event-by-event basis 
and, with sufficient statistics, yields a distribution of the scattering events 
in the total transverse momentum $\bK_\perp$.
Repeating this experiment at several values of $b$, one effectively scans the LG state 
and tracks the change of the $\bK_\perp$ distribution as a function of $b$.
In the absence of a phase vortex, we expect $\bK_\perp$ to be centered around zero
or shifted in the direction of $\bb_\perp$.
The presence of the phase factor $e^{i\ell \varphi_r}$, which exists only in the vicinity of a phase vortex,
unavoidably shifts the distribution in the direction orthogonal to $\bb_\perp$.
This shift grows as $b$ decreases, highlighting the superkick effect.

\begin{figure*}[t]
	\centering
	\includegraphics[width=\textwidth]{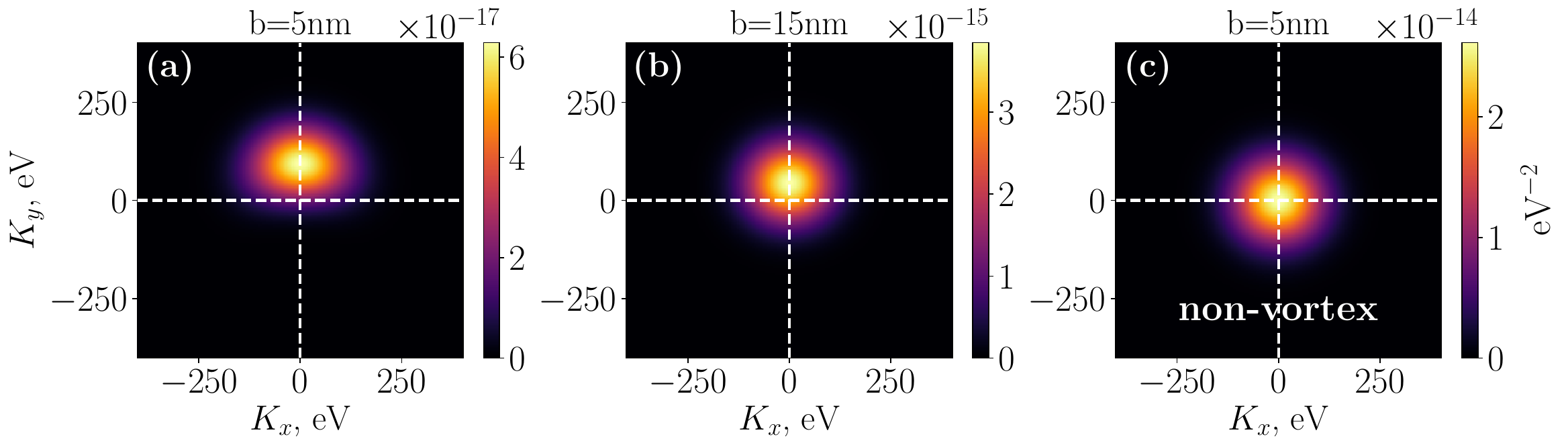}
	\caption{Differential scattering probability $w_K(\bK_\perp)$ for the Gaussian vs. LG state collision at $\bb_\perp = (b, 0)$ with $b = 5$~nm (a) and 15~nm (b) compared with the same distribution for non-vortex wave packet scattering (c).}
	\label{fig-b-dependence}
\end{figure*}

{\em Numerical results.}---To corroborate the above discussion, we undertook a numerical study 
of M\o{}ller scattering of an LG and a Gaussian wave packets using the exact expressions
for the wave functions and scattering amplitude given in the Supplemental material.
Our benchmark case corresponds to the following parameters of the initial electrons:
$p_{1z} = - p_{2z} = 10$~MeV, the total angular momentum of the LG electron $j_z = 7/2$,
both electrons unpolarized.
The wave packet extents are $\sigma_{1\perp} = 10$~nm, $\sigma_{2\perp} = 2$~nm,
so that $1/\sigma_{2\perp} = 100$~eV, $\sigma_{iz} = \sigma_{i\perp}/2$. 
The impact parameter is chosen along axis $x$: $\bb_\perp = (b, 0)$,
with values of $b$ ranging from zero to 40~nm. 

We integrate Eq.~\eqref{dsigma-WP} over the longitudinal momenta
and, keeping $\bK_\perp$ fixed, over $\bk_{3\perp}$
in the region $|\bk_{3\perp}| > k_{3 {\tiny \rm min}}$. This leads to the differential probability
\begin{equation}
	w_K = \frac{dW}{d^2K_\perp} = (2\pi)^2\int \frac{d^3k_{3}}{2E_3} \frac{d k_{4z}}{2E_4}\, |{\cal I}|^2\,.\label{I-K}
\end{equation}
Our task is to verify that $w_K$ indeed displays a shift
in the direction orthogonal to $\bb_\perp$.

In Fig.~\ref{fig-b-dependence}(a),(b), we show $w_K$ for $b = 5$ and 15~nm
and for the cut-off value $k_{3 {\tiny \rm min}} = 10$~keV,
which corresponds to the electron scattering angle of $10^{-3}$.
We observe a sizable shift of the intensity peak towards positive $K_y$ values,
which is orthogonal to $\bb_\perp$.
We checked that the dependence of $\lr{K_y}$ on $b$ follows 
the expected superkick curve discussed in \cite{barnett2013superkick,Ivanov:2022sco,Liu:2022nfq}.
To demonstrate that $\lr{\bK_\perp} \perp \bb_\perp$ is indeed an unmistakable feature 
of a phase vortex, we show in Fig.~\ref{fig-b-dependence}(c)
the same $w_K$ computed for a non-vortex state which lacks the vortex phase factor in momentum space.
As expected, off-axis collision of non-vortex states
does not generate the $\lr{\bK_\perp}$ component orthogonal to $\bb_\perp$.


{\em Feasibility study.}---The plots in Fig.~\ref{fig-b-dependence} 
correspond to the ideal situation of unlimited statistics, 
absolutely stable beams, and perfect final electron momentum reconstruction.
In a realistic experiment, these conditions will not be met, which requires us to study 
how uncertainties and limited statistics affect the visibility of the effect.

\begin{figure*}[t]
	\centering
	\includegraphics[width=\textwidth]{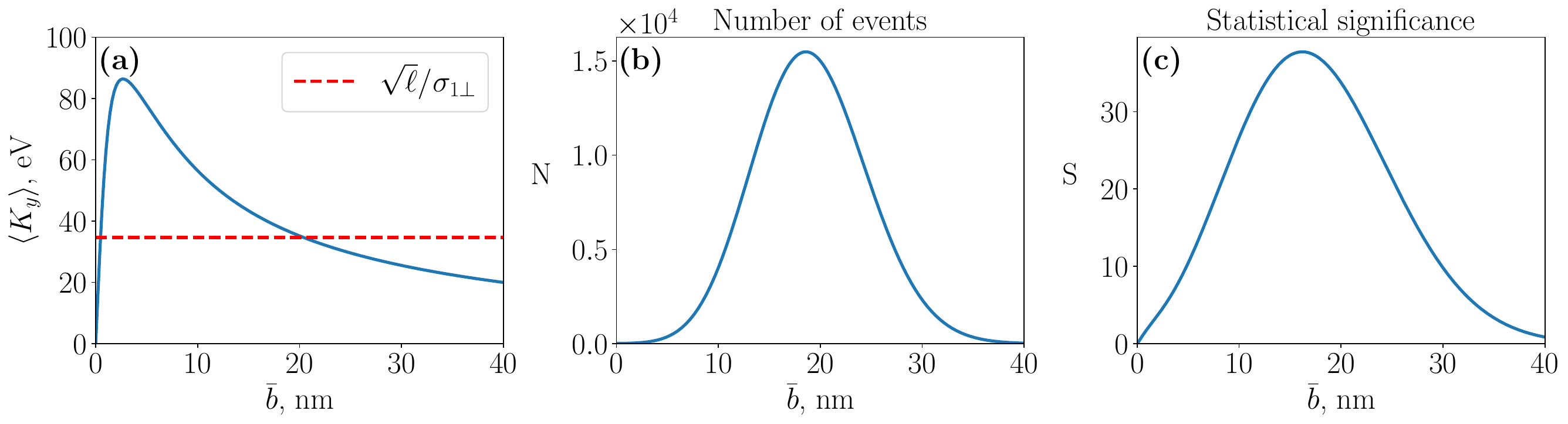}
	\caption{The value of $\lr{K_y}$ (a), the number of events (b), and the statistical significance of a non-zero $\lr{K_y}$ (c)
		as functions of the average impact parameter $\bar b$. 
		Transverse jitter and a finite momentum resolution are taken into account.}
	\label{fig-SS}
\end{figure*}

We consider the transverse localization parameters $\sigma_{1\perp}$ and $\sigma_{2\perp}$ realistic.
Much tighter focusing of vortex electrons, down to 0.1~nm, has already been demonstrated 
in electron microscopes \cite{Verbeeck:2011-atomic}.
However, bringing two wave packets in collision can lead to 
transverse jitter, event-by-event fluctuations of the impact parameter $\bb_\perp$. 
We allow it to fluctuate around $\lr{\bb_\perp} = (\bar b, 0)$ 
in an azimuthally symmetric range with radius $|\Delta \bb_\perp| = 2 \sigma_{2\perp}$,
as big as the Gaussian probe diameter. 
To account for it, we average $w_K$ over the above range of $\bb_\perp$
and obtain the smeared differential probability $\bar w_K = \lr{w_K}_b$. 
The average value of $\bK_\perp$ is now computed based on the smeared probability $\bar w_K$.
Longitudinal jitter is negligible
due to the large Rayleigh length: 
$L = \sigma_{2\perp}^2 |p_{2z}| \sim 200\,\mu$m. 

To estimate the expected number of events,
we integrate $\bar w_K$ with respect to $\bK_\perp$ and obtain the total scattering probability $W(\bar b)$.
At $b \to 0$, the two wave packets miss each other, and the probability is extremely small.
However, for $b \sim \sigma_{1\perp}$, it grows to ${\cal O}(10^{-10})$,
see details in the Supplemental material.
Multiplying by the total number $N_0$ of the electron pair collisions,
we expect to observe on average $N(\bar b) = N_0 W(\bar b)$ scattering events.
For definiteness, we consider electron beams at the average current 
of $16$~nA, which corresponds to $10^{11}$ electrons per second. 
At this current, sequential electrons in each beam are separated by $3$~mm,
large enough to neglect the potentially detrimental space-charge effect on vortex electrons.
An experiment running for $10^3$ seconds yields $N_0 = 10^{14}$ electron collision attempts;
at $b\sim \sigma_{1\perp}$, we expect to detect $\sim 10^4$ events.

For detecting the scattered electrons, we imagine, for reference, centimeter-scale forward region pixelized detectors installed 
in each direction 2~m downstream the collision point. 
To separate the two beamlines, collisions can be organized at a suitable non-zero crossing angle $\theta \not = 0$.
We checked through full numerical calculation that the effect remains well visible even for the large $\theta = \pi/4$.
Each detector can bear a 4~mm-wide central aperture
to let the electrons with $|\bk_{3\perp}| < k_{3 {\tiny \rm min}}$ pass. 
The pixel size of 20~$\mu$m translates into the angle uncertainty of $10^{-5}$
and the scattered electron transverse momentum uncertainty $\delta k_\perp = 100$~eV, 
which is comparable to the $\bar w_K$ shift in the $\bK_\perp$ plane.
Note that, with such momentum resolution, the detector would be unable 
to resolve the momentum-space ring of the LG wave packet by direct imaging. 
Yet, in elastic scattering, the tightly focused Gaussian wave packet amplifies the transverse scale of the effect 
and renders it visible.

Given all these imperfections and limited statistics, 
our goal is to detect with sufficient confidence
that $\lr{\bK_y}$ is indeed non-zero.
The figure-of-merit is the statistical significance which we define as
\begin{equation}
	S(\bar b) = \frac{\lr{\bK_y} \sqrt{N(\bar b)}}{\sqrt{\lr{\bK_y^2} + (\delta k_\perp)^2}}\,.\label{SS2}
\end{equation}
We consider the effect to be clearly visible if $S > 5$.

In Fig.~\ref{fig-SS}, we show $\lr{K_y}$, the number of scattering events $N(\bar b)$, and the statistical significance
of the non-zero $\lr{K_y}$ as we scan $\bar b$ across the vortex.
Although the superkick effect is expected to be the strongest at $b \ll \sigma_{1\perp}$,
the very low scattering probability and jitter make the non-zero $\lr{K_y}$ poorly visible.
However, for $b \sim \sigma_{1\perp}$, the statistical significance grows well above 5
and makes the effect clearly visible.


{\em Conclusions.}---In summary, collisions of high-energy vortex states 
will offer a wealth of information in atomic, nuclear, and particle physics. 
Their realization will critically depend on the unambiguous detection 
that a high-energy particle is indeed in a vortex state.
Existing low-energy diagnostic tools for vortex electrons are drastically limited 
in the MeV energy range, and other methods proposed in literature 
face serious challenges.
Here, we described a novel, superkick-based method to unmistakably detect the vortex state 
of a high-energy electron, which is free from the drawbacks of the other suggestions.
We proposed to perform elastic scattering of a compact Gaussian probe electron with 
a vortex electron state at a transverse impact parameter $\bb_\perp$,
detect the two scattered electrons using traditional detectors, 
measure their total transverse momentum $\bK_\perp$,
and observe that, on average, it is shifted orthogonally to $\bb_\perp$.
This feature can only be driven by the presence of a phase vortex and cannot be mimicked by non-vortex wave packets.
Numerical calculations taking into account experimental imperfections
confirm that the effect can be detected in a realistic experimental setting.

Although we performed our numerical calculation for 10~MeV vortex electrons, the effect has a weak sensitivity 
to energy and can be observed with the 300~keV vortex electrons already demonstrated in electron microscopes,
provided they can be brought in collision with a tightly focused non-vortex electron beam.
Such a proof-of-principle experiment will support the proposal and, at the same time, demonstrate
for the first time the reality of the superkick effect itself.
Concrete predictions for realistic experimental settings taking into account the parameters and limitations 
of the source and the detectors will be published elsewhere.
An extension of this idea to detecting high-energy vortex photons, ions, and hadrons is straightforward and
will also be elaborated in a future work.

\bigskip
{\em Acknowledgments.}---B.L. and I.P.I. thank the Shanghai Institute of Optics and Fine Mechanics for hospitality during their visit. S.L. and L.J. thank Sun Yat-sen University for hospitality during the workshop ``Vortex states in nuclear and particle physics''. S.L. and L.J. acknowledge the support by National Science Foundation of China (Grant No. 12388102), CAS Project for Young Scientists in Basic Research (Grant No. YSBR060), and the National Key R{\&}D Program of China (Grant No. 2022YFE0204800).

\bibliography{super-ee-arxiv-v2}
\bibliographystyle{apsrev4-2}

\bigskip

\section*{Supplemental Material}

\subsection{Wave packets of initial electrons}

Our calculations follow the scheme described in \cite{Liu:2022nfq},
which builds upon the general formalism of wave packet collisions \cite{Kotkin:1992bj,Karlovets:2020odl}.
Beginning with scalar particle collisions,
we describe the initial states by momentum space wave packets $\phi_1(\bk_1)$ and $\phi_2(\bk_2)$ normalized as
\begin{equation}
	\int \frac{d^3 k}{(2\pi)^3}\, \frac{1}{2E(k)}\, |\phi_i(\bk)|^2 = 1\,.
\end{equation}
The average momenta in each wave packets, $\bp_1 = \langle \bk_1\rangle$ and $\bp_2 = \langle \bk_2\rangle$, 
are antiparallel to each other and define the common axis $z$: $p_{1z} > 0$, $p_{2z} < 0$. 
We adopt the paraxial approximation, in which the typical transverse momenta are assumed to be much smaller 
that $p_{1z}$ and $|p_{2z}|$. Strictly speaking, a localized wave packet is non-monochromatic,
but in the paraxial approximation the energy distribution peaks around $\varepsilon_1 = \sqrt{p_1^2 + m^2}$
with an energy spread that is negligible for our purposes.
The coordinate wave functions are defined by
\begin{equation}
	\psi(\br,t) = \int \frac{d^3k}{(2\pi)^3\sqrt{2E(k)}}\,\phi(\bk)\, e^{i\bk\br - iE(\bk)t}\label{psi-def}
\end{equation}
and are normalized as $\int d^3r\, |\psi(\br,t)|^2 = 1$.

For the scalar vortex state, we use the relativistic LG principal mode:
\begin{eqnarray}
	&&\phi_1(\bk_1) = (4\pi)^{3/4}\sigma_{1\perp}\sqrt{\sigma_{1z}} \sqrt{2E_1}\, \frac{(\sigma_{1\perp}k_{1\perp})^{\ell}}{\sqrt{\ell!}}\nonumber\\
	&&\times 
	\exp\left[ -\frac{k_{1\perp}^2 \sigma_{1\perp}^2}{2}-\frac{(k_{1z}-p_{1z})^2 \sigma_{1z}^2}{2} +i\ell\varphi_k\right] \,.
	\label{LG-full-def}
\end{eqnarray}
Here, $\sigma_{1\perp}$ and $\sigma_{1z}$ are the transverse and longitudinal spatial extents of the coordinate space wave function,
which are taken as independent parameters. We choose $\sigma_{1z}$ to be of the same order as $\sigma_{1\perp}$
to guarantee that the wave packets do not significantly spread during the overlap time; this regime was called in \cite{Ivanov:2022sco}
the impulse approximation.
Using the LG modes with the radial quantum number $n > 0$ or the so-called elegant LG wave packets 
as defined in \cite{Sheremet:2024jky} would also be possible, but it is not essential for presenting our main idea.
The Gaussian state can be obtained from the above formula by setting $\ell = 0$: 
\begin{eqnarray}
	&&\phi_2(\bk_2)=(4\pi)^{3/4}\sigma_{2\perp}\sqrt{\sigma_{2z}} \sqrt{2E_2}\,\, e^{-i b_z k_{2z} + i \tau E_2}   \nonumber\\
	&&\times \exp\left[-\frac{k_{2\perp}^2 \sigma_{2\perp}^2}{2}-\frac{(k_{2z}-p_{2z})^2 \sigma_{2z}^2}{2} 
	-i \bb_\perp \bk_{2\perp}\right]\,,
\end{eqnarray}
with parameters $\sigma_{2\perp}$ and $\sigma_{2z}$.
Here, we also take into account the possibility that the two wave packets may be shifted 
with respect to each other in three different ways.
The impact parameter $\bb_\perp$ defines the transverse offset between their axes,
$b_z$ defines the longitudinal distance between their focal planes, 
and $\tau$ characterizes the time difference between the instants of their maximal focusing. 
As argued in \cite{Ivanov:2022sco,Liu:2022nfq}, in the paraxial approximation, $b_z$ and $\tau$ play inessential roles 
due to the large Rayleigh length $L_R = \sigma_{2\perp}^2 |p_{2z}|$.
Thus, we can safely set $b_z$ and $\tau$ to zero.

To take into account that electrons are spin-1/2 particles, we describe the internal degrees of freedom of the plane-wave electron 
with the four-momentum $k^\mu$ and helicity $\lambda$ with the bispinor $u_{k\lambda}$: 
\begin{equation}
	u_{k\lambda} = \doublet{\sqrt{E+m_e}\,w^{(\lambda)}}{2 \lambda \sqrt{E-m_e}\,w^{(\lambda)}}\,. \label{PW-bispinors}
\end{equation}
Our choice of the spinors $w^{(\lambda)}$ is slightly different for the first and the second electron in M\o{}ller scattering.
For the first electron in M\o{}ller scattering, the one with $k_{1z} > 0$, the spinors are
\begin{equation}
	w^{(+1/2)} = \doublet{c_1}{s_1\, e^{i\varphi_1}},
	\quad w^{(-1/2)} = \doublet{-s_1\, e^{-i\varphi_1}}{c_1}\,,\label{PWspinors-1}
\end{equation}
while the second, counter-propagating electron with $k_{2z} < 0$, is described with
\begin{equation}
	w^{(+1/2)} = \doublet{c_2 e^{-i\varphi_2}}{s_2},
	\quad w^{(-1/2)} = \doublet{-s_2}{c_2\,e^{i\varphi_2}}\,.\label{PWspinors-2}
\end{equation}
Here, $c_i \equiv \cos(\theta_i/2)$, $s_i \equiv \sin(\theta_i/2)$, with $\theta_i$ and $\varphi_i$ 
being the polar and azimuthal angles of $\bk_i$.
This choice possesses a well-define forward limit for the first electron $\theta_1 \to 0$,
when $s_1 \to 0$, and a well-define backward limit for the second $\theta_2 \to \pi$, when $c_2 \to 0$.
Similar conventions are used for the two final electrons in small-angle scattering.
Note that, for non-zero $\theta_i$, these bispinors are not eigenstates of the operators
of the $z$-projections of spin or OAM. 
However, they are eigenstates of the $z$-projection of the total angular momentum operator $\hat j_z = \hat s_z + \hat \ell_z$
with the eigenvalues $j_z = \lambda_1$ for the first electron and $j_z = -\lambda_2$ for the second electron. 

We define the initial electron wave packets in momentum space as $\phi_i(\bk_i)u_{k_i\lambda_i}$.
The resulting LG state of the first electron is not an eigenstate of the OAM or spin $z$-projections,
but it is still an eigenstate of $\hat j_z$ with the eigenvalue of $\ell + \lambda_1$, where $\ell$ is the parameter
of the LG state in Eq.~\eqref{LG-full-def}.
Nevertheless, in the paraxial approximation, $\lr{\hat \ell_z} \approx \ell$, and although the OAM
and the spin degrees of freedom are entangled due to the unavoidable spin-orbital interactions,
one can conduct the calculations keeping in mind that the OAM and spin are approximately conserved.

\subsection{M\o{}ller scattering of wave packets}

In the main text, we defined the differential probability of the elastic scattering $dW$, which is the main observable in
localized wave packet scattering. If needed, one can define the generalized cross section as $d\sigma = dW/L$,
where the luminosity function, or the flux, for a general two wave packet collision can be found in \cite{Karlovets:2020odl}. 
Within the paraxial approximation, the relative velocity $|v_1-v_2|$ can be computed 
via the average momenta of the two wave packets $v_i = p_{iz}/\varepsilon_i$ (with $v_1>0$ and $v_2<0$),
which simplifies the expression and allows us to represent the luminosity function as a space-time overlap 
of the two colliding wave packets:
\begin{equation}
	L = |v_1 - v_2| \int d^3 r\, dt\, |\psi_1(\br,t)|^2 |\psi_2(\br,t)|^2\,.\label{lumi}
\end{equation}
This factor takes care of the correct normalization which is especially important when two the colliding wave packets overlap only partially.
We stress once again that the primary observable is the scattering probability itself, which we analyze in the main text.

Next, the amplitude of the wave packet scattering into final plane waves is expressed 
via an appropriately weighted plane wave helicity amplitude ${\cal M}$.
In the Born approximation, the plane-wave $ee\to e'e'$ helicity amplitude can be found, for example, in \cite{Landau4}
and is given by
\begin{equation}
	{\cal M} = e^2\left({\bar u_3 \gamma^\mu u_1\, \bar u_4 \gamma_\mu u_2 \over t}
	- {\bar u_4 \gamma^\mu u_1\, \bar u_3 \gamma_\mu u_2 \over u}\right).
\end{equation}
The exact expressions for the helicity amplitudes $\lambda_1 \lambda_2 \to \lambda_3\lambda_4$
in the generic kinematics were given in \cite{Ivanov:2016oue} and are used in our numerical calculations. 
We follow the prescription of \cite{Serbo:2015kia,Ivanov:2016oue}
to define the unpolarized vortex electron as an equal mixture of positive and negative helicity states
with the same value of the conserved quantity $j_z$, that is, 
the LG states with $\ell = j_z - 1/2$, $\lambda_1 = +1/2$ and $\ell = j_z + 1/2$, $\lambda_1 = -1/2$.
The key conclusions of our paper would remain the same if an alternative definition 
of the unpolarized LG electron based on the same value of the parameter $\ell$ were used.

\subsection{Estimating the scattering probability}

The visibility of the effect depends on the expected number of events. 
To get a better insight into the results of the numerical calculations shown in the main text, 
we describe here an analytic order of magnitude estimate of this quantity. 

As shown in \cite{Karlovets:2020odl}
and explicitly verified in the superkick-related situation in \cite{Liu:2022nfq}, 
the integrated cross section of wave packet scattering well above the threshold is very close to the plane wave cross section $\sigma_{PW}$.
The novel effects of the vortex state collisions appear primarily in the differential cross sections, not in the total ones.
The interaction probability can be well approximated by $W = \sigma_{PW}\cdot L$, with the luminosity function
given in Eq.~\eqref{lumi}. The elastic scattering cross section integrated over $|k_{3\perp}| > k_{3 {\tiny \rm min}}$ is
\begin{equation}
	\sigma_{PW} = \frac{4\pi\alpha_{em}^2}{k_{3 {\tiny \rm min}}^2} \approx 
	\frac{2\times 10^{-5}\,\mbox{nm}^2}{(k_{3 {\tiny \rm min}}\, [\mbox{keV}])^2}\,.
\end{equation}
The analytical expression for $L$ was derived in \cite{Liu:2022nfq} and, for $\sigma_{2\perp} \ll \sigma_{1\perp}$,
it can be approximated by
\begin{equation}
	L \approx \frac{1}{\pi\, \ell!\, \sigma_{1\perp}^2} \left(\frac{b^2}{\sigma_{S1\perp}^2}\right)^\ell e^{-b^2/\sigma_{1\perp}^2}\,.
\end{equation}
This expression is maximal at $b = \sqrt{\ell} \sigma_{1\perp}$, and for moderately large $\ell$ it yields $L \sim 0.1/\sigma_{1\perp}^2$.
As a result, the order of magnitude estimate of the maximal scattering probability is
\begin{equation}
	W \sim \frac{10^{-6}}{\bigl(k_{3 {\tiny \rm min}}[\mbox{keV}] \cdot \sigma_{1\perp}[\mbox{nm}]\bigr)^2}\,.
	\label{estimate}
\end{equation} 
For $k_{3 {\tiny \rm min}} = 10$~keV and $\sigma_{1\perp} = 10$~nm, it gives $W \sim 10^{-10}$ 
and, with sufficient current and experiment running time,
leads to thousands of detected events. 
A stronger focusing of the LG wave packet would boost
the probability but, at the same time, would require us to increase the lower limit $k_{3 {\tiny \rm min}}$
is order to be able to safely distinguish scattered and non-scattered states.
If $b\ll \sigma_{1\perp}$, the luminosity function decreases rapidly, and the number of events may drop below one.

\end{document}